# Hybridization and Dehybridization of Plasmonic Modes

Artur Movsesyan,*,| Alina Muravitskaya,*,# Marion Castilla, | Sergei Kostcheev, | Julien Proust, |
Jérôme Plain, | Anne-Laure Baudrion, | Rémi Vincent, | and Pierre-Michel Adam|

| Laboratory Light, nanomaterials and nanotechnologies - L2n, University of Technology of Troyes and CNRS ERL 7004, 12 rue Marie Curie, 10000 Troyes, France

# B.I. Stepanov Institute of Physics, National Academy of Sciences of Belarus, Nezavisimosti 68-2, Minsk, 220072, Belarus

E-mail: movsesyan@gmail.com; alina.muravitskaya@gmail.com

## ABSTRACT

The plasmon resonances (modes) of a metal nanostructure can be defined as a dipole, a quadrupole, or high-order modes depending on the surface charge distribution induced by the incident field. In a non-symmetrical environment or clusters, the modes can hybridize and exhibit different behavior and properties. In this work, we study experimentally and numerically the substrate-induced hybridization of plasmonic modes of a silver nanocylinder. The applications of plasmonic nanoparticles such as refractive index sensing and enhanced spectroscopies often rely on the sustained mode spectral position and specific spatial near-field distribution. However, we show that the implementation of such plasmonic nanoparticles in a sensing system can result in a change of the modes nature, its hybridization or dehybridization. These changes are not clearly pronounced in the far-field spectra and then may result in unexpected modifications of the sensor behavior. We show that the hybridization between the dipolar and quadrupolar modes of the plasmonic nanoparticle on the substrate results from quadrupolar mode excitation due to the superposition of the reflected and incoming light and, then, depends on the reflection of the substrate. The existence of the hybridized modes strongly depends on the surrounding environment of nanoparticles, and after the deposition of the nanometric polymer layer on top of the nanoparticle the hybridized modes vanish and are replaced by uncoupled multipolar resonances.

## INTRODUCTION

The optical properties of metal nanoparticles associated to excitations of the localized surface plasmons resonances (LSPRs) are important in various research areas such as surface-enhanced spectroscopies,[1-4] nanophotonics,[5] biomolecular and chemical sensing,[6-9] nanolasing,[10] photocatalysis,[11,12] photothermal therapy.[11] LSPRs are defined as a collective oscillation of conduction band electrons of a metal nano-object, driven by an external electric field. These resonances (modes) are sensitive to different factors such as the size, shape, morphology of nanostructures, the chosen metal material, and the refractive index (RI) of the surrounding medium.[13,14] There are many studies about spectral tunability of the plasmonic modes and applications of this phenomenon such as high-performance refractive index sensors based on the LSPR spectral shift.[6,13]

For applications, besides the spectral position of plasmonic resonances, the spatial distribution of the intensive electromagnetic field near the nanoparticle is important. The confinement of the field depends on the excited plasmonic mode nature and the shape of the nanoparticle. A plasmonic nano-object may exhibit dipolar mode excitation as well as high-order resonances if the necessary excitation conditions are fulfilled.[15-18] The plasmonic modes can be bright or dark depending on their optical behavior.[19] The modes, which couple with incident light and scatter, are bright. The dark modes have an effective zero dipole moment and do not radiate in contrast to the bright modes. The spatial distribution of the electromagnetic field for the dipolar mode of the nanostructure usually features two symmetrical lobes coinciding with the incident field polarization direction. Interestingly, the nanostructures placed on substrates with higher refractive index than air mostly sustain plasmonic mode with a near-field distribution confined to the substrate.[20] In some cases, this redistribution of the field also goes along with the appearance of an additional mode in the spectrum, which has an opposite field distribution with the highest field intensity at top of the nanostructure.[16,21-27] The substrate influence on the plasmonic modes spectral positions and optical behavior was studied in two directions: to eliminate the confinement of the field near the substrate because it reduces the sensitivity of the RI plasmonic sensors,[20,28-30] or to design plasmonic nanostructures with a controllable excitation of the mode with high electric field confinement on the nanostructure's top surface.[23,31] In the latter case, the authors supposed that the spatial distribution of the intensive electromagnetic field near the nanoparticle should be beneficial for numerous applications.[23,24,31] For instance, in bio-chemical sensing compounds are usually deposited on top of the nanoparticles, and a plasmonic mode confined to the air side should provide a stronger interaction with analytes than the mode confined to the glass.

Sherry et al. showed experimentally and numerically the existence of two bright modes for a single silver nanocube near the substrate.[24] However, on the experimental scattering spectrum, two prominent modes appeared only for a single nanocube. The ensemble measurements showed only one extinction peak. The second one disappeared due to the inter-particle coupling and averaging effects. Later on, Zhang et al. proposed a model to explain the origin of these modes based on the hybridization theory.[23] They showed that in a non-symmetrical environment under certain conditions, a coupling between a dark quadrupolar and a bright dipolar plasmon modes happens. This process depends on the relative spectral positions of the modes. The asymmetric environment (substrate) mainly causes two phenomena: (i) a possible interference between incoming and scattered fields; (ii) a spectral shift of plasmonic modes due to the dielectric screening. Later Lermé et al. modeled the separation of these two phenomena and their impact on the hybridization of plasmonic modes of a silver nanosphere using analytical and numerical methods.[22] The results show that one of the key conditions for the hybridization of plasmonic modes is the interference of the incoming and scattered fields. Several authors proposed to use this hybridized resonance for sensing.[23] Although the proposed nanostructures and model look promising in applications, some of the papers reported obstacles in observing well-determined calibrating curves for the modes of the nanoparticles on the substrate[30,31] which can result from the change of the system symmetry during the increasing or decreasing of the RI of the superstrate.

In this work we propose an experimental and numerical study of the hybridized plasmonic modes of silver nanocylinders. We consider the possibility and feasibility of using these hybridized modes in applications. Firstly, we discuss the origin and the nature of the two dominant hybridized modes arising from the interaction of the substrate and the nano-object. The modification of the hybridized modes due to changes of the superstrate refractive index is displayed. We also explain in details the hybridization and dehybridization of plasmonic modes excited in the nanoparticles near the substrate. The importance of the reflectivity of the substrate on the hybridization is also revealed. We show that it may limit the use of the hybridized modes in sensing and enhanced spectroscopies.

METHODS

The numerical modeling is performed with the help of commercial software FDTD solutions (Lumerical) based on finite-difference time-domain method. For a planewave illumination we use TFSF (total field scattered field) source.[32] The discretization mesh is 0.5 nm. As boundaries perfect matching layers (PMLs) are used.[33] For numerical simulation we use the experimentally measured silver dielectric permittivity data.[34] The glass (fused silica) substrate refractive index is 1.47.

The following formalism was used to calculate the distribution of charges:

$$\rho(r) = \epsilon_0 \nabla \cdot E$$

Where $\rho(r)$ is the charge density at $r$ position, $\epsilon_0$ is the permittivity of free space, and $\nabla \cdot E$ is the divergence of the electric field.

The silver nanocylinders (SNCs) samples were fabricated by electron beam lithography. We produced a disordered arrays to prevent the inter-particle coupling and diffraction effects of the nanoparticles arrays. The SNCs diameters were 70 nm, 80 nm and 90 nm. The height was 50 nm. The distance between the SNCs is from three to six times their diameter. The arrays measures $30 \times 30\ \mu m^2$. We used 3 nm adhesion layer of titanium oxide in order to assure good bonding between the substrate and the nanoparticles. Optical spectroscopy is done with a standard homemade microscope spectrometer using a collection lens (numerical aperture 0.2), equipped with a UV-visible lamp source, a fiber-coupled spectrometer (Ocean Optics USB2000+). Optical spectra are acquired in transmission (T), normalized by lamp signal (T_0), and plotted as $1 - T/T_0$. It measures the fraction of the light which is absorbed and scattered by the nanoparticles. The nanoparticles sizes are measured with the help of a scanning electron microscopy (Supporting Information Fig. S1). The nanoparticles of the smaller size (d=40 nm, h=30 nm) were produced and characterized by the same procedure. After the optical measurements the ~10 nm layer of PMMA (molar mass 25 kg/mol, concentration 3 g/L) was spin-coated on the sample to study the effect of the dehybridization. Then, the optical measurements were repeated.

RESULTS AND DISCUSSION

In Fig. 1 we show the calculated extinction spectra of a silver nanocylinder (SNC) in air and placed at different distances from a glass substrate. The extinction spectrum for the SNC in air shows a single peak (403 nm) and a shoulder (350 nm), while the extinction spectrum for the nanoparticle placed on the substrate shows two peaks (465 nm and 380 nm) and a shoulder (347 nm). These three plasmonic modes are marked on Fig. 1a as mode (i), mode (ii), mode (iii). One can see that while approaching the SNC to the substrate, the mode (i) mainly experiences a redshift of about 60 nm. The electric field map for this mode (Fig. 1b) shows the highest near-field intensity localized near the substrate interface. The charge distribution calculations confirm that the charges mostly concentrate near the substrate.

The electric field map of mode (ii) is depicted in the Fig. 1c. The main confinement is on the top edge of the SNC with the corresponding surface charge distribution. One may see that this mode has almost a dipolar kind of charge distribution even if charges are mostly localized on the top edges of the SNC. When the SNC approaches the substrate, the interactions with the glass do not change the spectral position of the mode (ii). It just appears more distinctively, mainly due to the redshift of the mode (i). Two similar modes were described in Ref. 23 as a result of the coupling and hybridization of the plasmon modes of a metal nanocube near the substrate due to the symmetry breaking of the system. Although this process was discussed in previous works, the mechanism of such hybridization is not fully understood. The quadrupolar mode in a nanocylinder is dark and, as such, is not excited directly by the incident light. The quadrupolar mode may become bright due to the retardation effect of the field if the nanoparticle size is comparable to the excitation wavelength. However, the hybridization effect near the substrate was observed for relatively small nanoparticles also.[22,23]

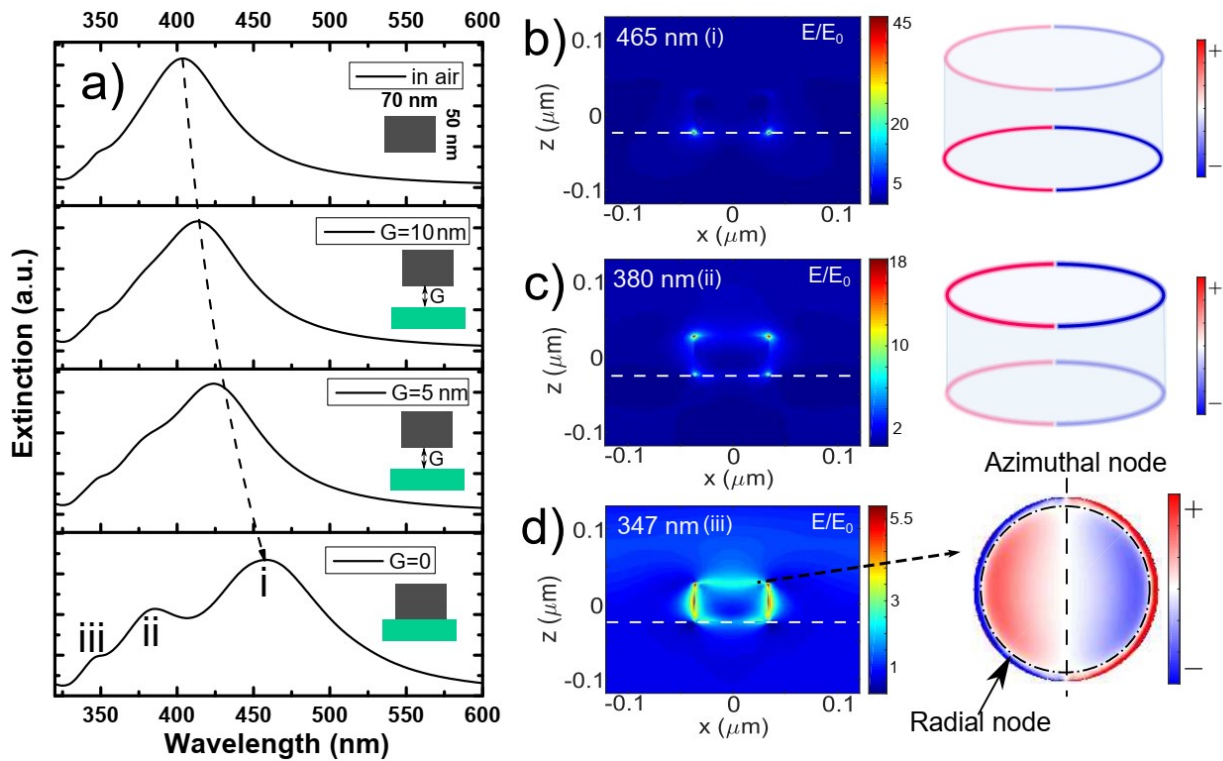

Figure 1: (a) Calculated extinction spectra of a silver nanocylinder in air and approaching the glass substrate on different distances (the diameter of SNC is 70 nm and the height is 50 nm). (b-d)

Electric field maps: XZ cut profiles in the center of nanocylinder on the glass for 465 nm excitation (Mode i) (b), 380 nm (Mode ii) (c), 347 nm (Mode iii) (d). White lines show the substrate interface. Near the electric field maps, there are the schemes representing calculated charges distribution of corresponding wavelengths. The vertical dash line shows the azimuthal node and the circular dash-dotted line corresponds to the radial node.

In our case, the 50 nm height SNC does not show a clear excited quadrupolar mode in the air (Fig. 1a). Thus, we may not attribute its excitation to the field retardation effect when the nanoparticle is on the glass surface. To understand why the quadrupolar mode is excited and may couple with the dipolar mode, we consider what physical effect happens when the nanoparticle approaches the substrate. Near the substrate, both incoming and reflected plane waves interact with the nanoparticle. Then, the inhomogeneity of the field, resulting from the interference between the incident field and reflected scattered field, changes the modes excitation. However, the scattered field includes radiative and evanescent components, which both contribute to the observed effect. The reflected wave from the substrate has a $\pi$ phase difference with the impinging wave. Also, for such small distances between a nano-object and substrate, the interactions of the evanescent fields of nanocylinder and the substrate may play a more significant role then the interference between the incoming and reflected field.

In order to show these substrate effects, we present a numerical simulation for a nanocylinder of similar dimensions, broken down into two parts (bottom and top parts). These two nanocylinders are used as illustration (probe) of the local field excitation and the dipole moments, rather than an equivalence to the thick nanocylinder. We break down the 50 nm height SNC in two nanocylinders of 10 nm height. The air gap between the two nanocylinders is 30 nm. In Fig. 2a we show the electric field vectors for the two nanoparticles when they are suspended in the air. One may note that both nanoparticles show dipole moments oriented in the same direction. The surface charge distributions are the same for both nanocylinders. Fig. 2b shows the dipole moments direction of the two nanoparticles close to the substrate. Herein, we see that the electric field vectors tend to opposite directions for the top and bottom nanoparticles, and they exhibit counter dipole moments. Indeed, this change happens gradually with the removal from the substrate. The animation in SI shows how the dipole moment orientation of the bottom nanocylinder rotates from in-phase oscillation with the upper nanocylinder to the out-of-phase depending on the distance between nanocylinders and the substrate. With the increase of the distance between two nanocylinder and the substrate from 10 nm to 20 nm the orientations of the dipole moments become the same for both nanocylinders, while for the small distances the dipole moments have opposite direction.

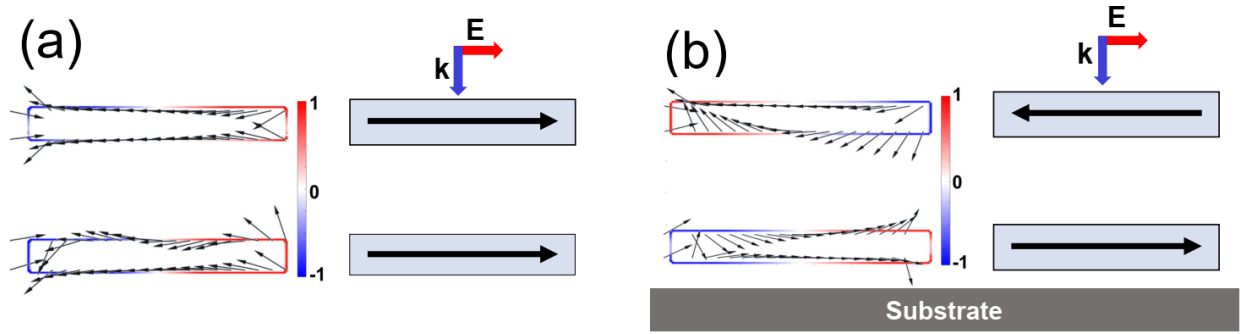

Figure 2: The model and calculations of the electric field vectors and charge distributions for two nanoparticles of 10 nm height separated by 30 nm gap placed in the air (a) and 10 nm above the glass substrate (b) in resonance conditions. The arrows on the right side of (a) and (b) shows the direction of electric displacement according to the computed electric field vectors. The arrows are laid on the surface charge distributions of the nanocylinders.

In Fig. 3 we present this hybridization model depicturing these modes as a sum and the difference of the dipolar and quadrupolar modes. One may observe that the antibonding mode (D-Q) has charge accumulation like mode (ii) and that the bonding mode (D+Q) has a similarity with mode (i).

It is important to note that when we approach the substrate, the effective refractive index around the SNC changes. In this manner both modes D and Q should be redshifted. According to the hybridization model the mode (i) is the sum of the D and Q modes, then the displacement of mode (i) should be a sum of the spectral shifts of modes D and Q. On the contrary, as the mode (ii) is the difference between the D and Q modes, then its dependence on the surrounding RI is defined by the difference between the modes D and Q spectral shifts. One can see that mode (ii) experience a very small red-shift because of the effective refractive index change (Fig. 1a).

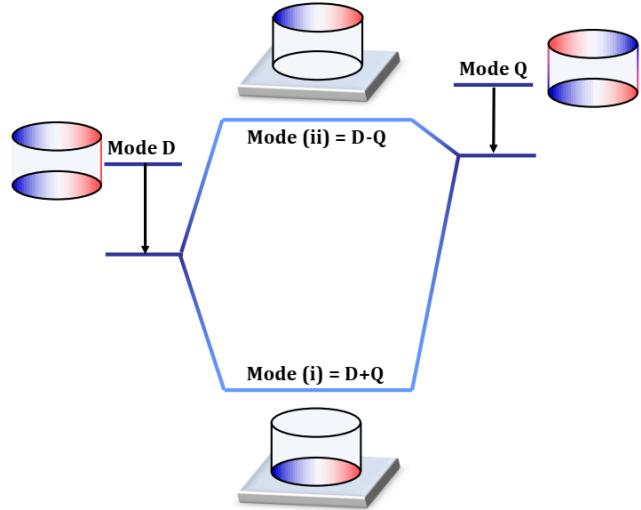

Figure 3: Model of the plasmonic modes hybridization in a nanocylinder due to an asymmetric environment. Mode (Q) and mode (D) are quadrupolar and dipolar primitive plasmon modes of a nanocylinder. Near the substrate, these two modes redshift which is shown by black arrows. Mode (ii) and mode (i) are hybridized antibonding and bonding modes.

The third mode, present in the spectrum of the SNC (Fig. 1a), is named mode (iii). Its spectral position is almost unchangeable for different distances from the substrate. The mode (iii) is not affected by the interaction with the dielectric substrate in contrast to mode (i). To understand the origin of this mode we show the electric field map at 345 nm excitation in Fig. 1d. The electric field is confined to the particle in all planes of observation. The calculated surface charges distribution for the top plane of the nanoparticle is presented. The mode (iii) shows all features of the pseudo breathing mode (1,1),[15,35,36] where there is a radial and azimuthal modes admixture. Moreover, it has a surface charges distribution similar to the resonant cavity modes of patch nano-antennas.[36,37]

The Fig. 4a shows the experimentally obtained extinction spectrum of the disordered SNCs (the diameter is 70 nm and the height is 50 nm) array. One may notice two defined peaks at 490 nm and 358 nm in contrast to the three peaks pronounced in the simulations (Fig. 1a). Also, the left side of the main peak is not symmetrical and has a shoulder located around 425 nm. One can assume that it happens due to the differences between the simulation model and real experimental nanoparticles and conditions. The SNC produced by electron beam lithography does not have the exact profile such as perfect cylinder used as a model in the numerical simulations. The real nanocylinders profile has a curvature in the top plane and a small bending angle of the nanocylinder side walls (inset of Fig. 4b). To compare more carefully the experimental results with simulations, we show in Fig. 4b the numerically simulated extinction spectrum of SNC considering the cylinder profile and the presence of the adhesion layer. One may observe a good match between experimental and simulation spectra. The similar behavior was observed for the nanocylinders of diameter 80 nm and 90 nm (Supporting Information Fig. S1).

The field maps (Fig. 4c-f) allow us to distinguish the new spectral positions of the modes and name them in the spectrum. The mode (i) has the strong red-shift compared to the simulations for the nanocylinder on the glass substrate (Fig. 1a). This can be explained by the influence of the higher refractive index of titanium oxide than glass. Especially this mode is basically localized near the substrate interface, which increases the impact of the adhesion layer.

The mode corresponding to the shoulder at 425 nm has high confinement of the electric field in the adhesion layer (Fig. 4e), which is different from the system without the TiO$_2$ layer (Fig. 1a). Then, we assume that this mode is attributed to the titanium oxide layer. The layer of titanium oxide is very thin and not chemically uniform since it was produced by electron-beam evaporation. This is the reason why we do not see a prominent shoulder in the experimental spectrum, unlike in the simulations.

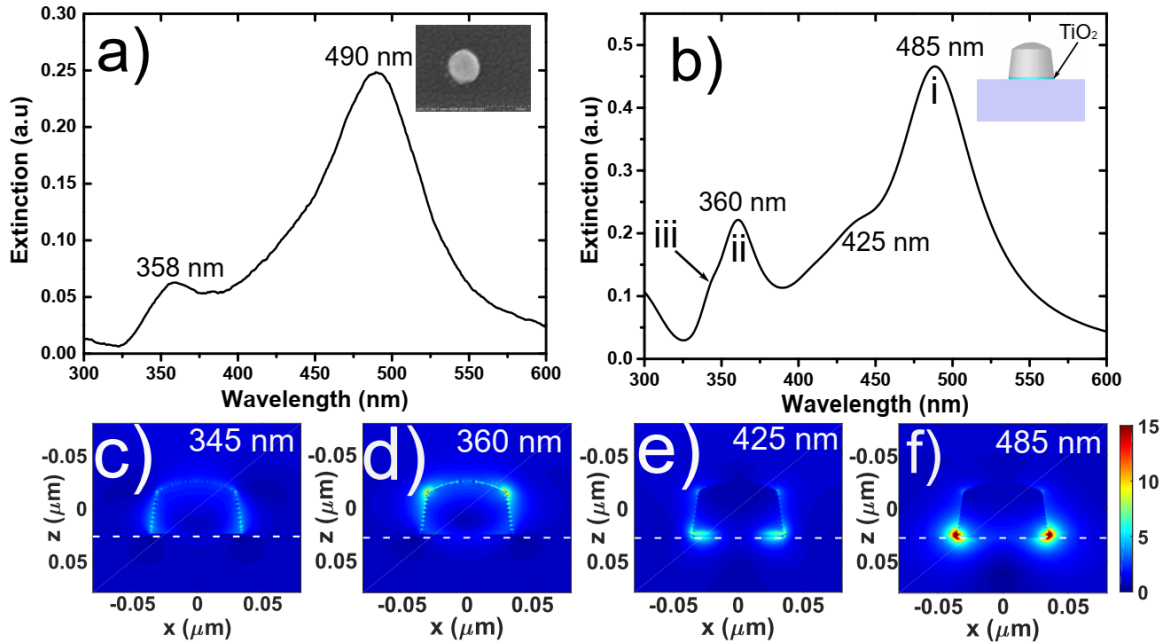

Figure 4: (a) Experimental extinction spectrum of a SNCs, the diameter is 70 nm and the height is 50 nm, (b) Corresponding numerical simulation. (c) Normalized electric field maps: XZ cut profile in the center of nanocylinder. Illumination is along z-axis from top side and polarized along x axis.

White dashed lines show the substrate surface. EF map for 345 nm excitation (Mode iii), (d) 360 nm (Mode ii), (d) 425 nm (magnetic mode), (f) 485 nm (mode i). All the maps were saturated to $|E/E_0| = 15$.

The modification of the nanoparticle shape (decrease of the size of the top plane) results in a change of the spectral position of the mode (ii), and it overlaps with mode (iii) more. In fact, for the mode (i), which is basically localized near the substrate interface, the red-shift occurs due to the titanium oxide layer. On the contrary, the mode (ii) shifts due to the change of the nanoparticle top part. When in the simulation the top plane diameter is reduced, the mode (ii) spectrally blue-shifts. Even though the mode (ii) is modified after SNC profile changes, the electric field map (Fig. 4d) shows confinement of energy on the top plane. The mode (ii) and mode (iii) almost spectrally overlap, and it is hard to observe the mode (iii) in experiments as well as in simulations.

As we mentioned above, the hybridized plasmonic mode (ii) is highly interesting for applications in sensing and enhanced spectroscopies. Since the electric field distribution of the conventional dipolar mode is concentrated on the substrate side, it is not reachable for the species of interest like emitters or molecules which may cover the plasmonic nanoparticles. Whilst the hybridized mode (ii) is confined in the top part of the nanoparticle exposing the hot spots to the species of interest. We then check the possibility and compatibility of the mode (ii) for future applications in enhanced spectroscopies or sensing. In Fig. 5 we depict a simulated extinction spectrum (blue curve) of 50 nm height and 70 nm-diameter SNC covered with a layer of the dielectric of 1.5 refractive index. It represents an usual polymer layer that is often used to deposit the emitters or molecules of interest.[6,38,39] In comparison with the uncovered SNC on the glass substate, herein we observe four modes denoted α, β, γ and δ. From the first view we note that mode (γ) is at the spectral position (around 380 nm) of the plasmonic mode (ii), which we observed previously in Fig. 1a. Moreover, the spectral position of the mode (δ) matches that of the mode (iii).

The comparison of the absorption (red dashed line) and the scattering spectrum (black dotted line) shows that the mode (β) has a well-defined peak in absorption and an insignificant shoulder in scattering, which may announce the presence of a quasi-dark mode. In order to precisely determine the nature of all excited modes we calculated their charges distributions. The mode (α) shows a dipolar distribution. Mode (β) shows an out-of-plane quadrupolar mode distribution. The neighboring mode (γ) has a mixed character between intensive out-of-plane hexapolar mode and less intensive mode (δ). These modes are spectrally close to each other, which results in an influence of mode (δ) on the hexapolar charge distribution. The mode (δ) is similar to mode (iii). One may clearly notice that there is no mode similar to mode (ii). We have new excited modes like out-of-plane quadrupolar (β) and hexapolar modes (γ). When we increase the refractive index around the nanoparticle, the wavelength of the incident light is reduced with respect to air values, then the ratio of the nanoparticle height to the wavelength of incident light increases. Thus, the higher phase variation inside the nanocylinder occurs, and the multipole modes like the quadrupole and the hexapole are excited due to the retardation effect but not due to the hybridization process. As we have shown previously the hybridization takes place due to interaction with a reflection wave coming from the substrate, then when the SNC is partially covered with a material with a similar RI to the substrate, the hybridization does not happen.

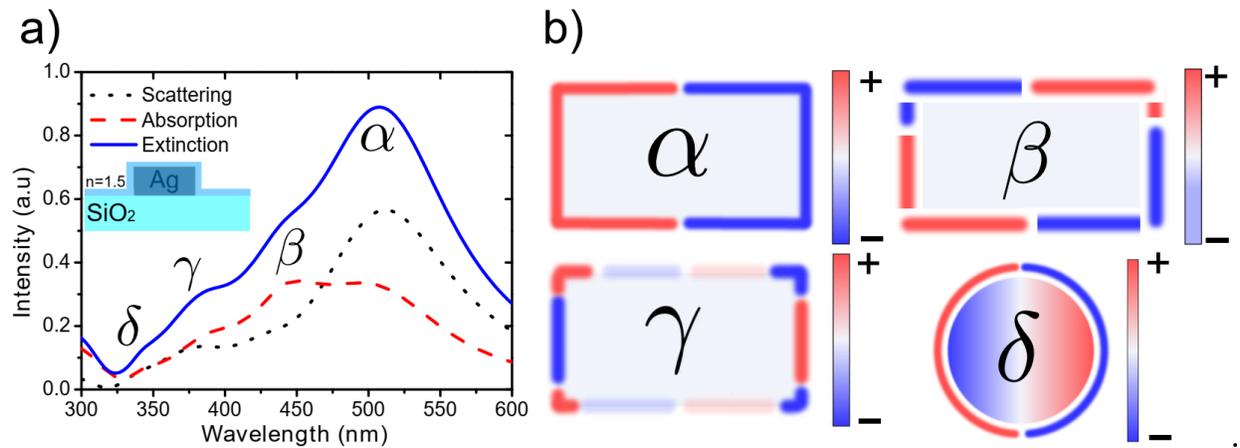

Figure 5: (a) Simulated extinction, absorption and scattering spectrum of a SNC (70 nm-diameter, 50 nm-height) covered by 10 nm-height layer of a refractive index of 1.5. (b) Visualization of charge distributions from computed data (SI, Fig. S2) for the peak position of α, β, γ (XZ profiles) and δ (XY profile) modes.

Then, if we cover the SNC (d= 70 nm, h= 50 nm) lying on the substrate, with a layer having a refractive index comparable to that of the substrate we can excite quadrupolar and hexapolar modes due to the retardation effect, although the hybridized modes vanish. Moreover, at the spectral position of mode (ii) we have another multipole (hexapole, mode γ), which can create confusion for sensing. It is important to note that when we deposit a layer of refractive index close to the index of glass, the environment of the nanoparticle becomes more homogeneous. This equalization of the local refractive index results in the redistribution of the field and reduction of the symmetry breaking effects. The maximum of the electric field relocates from the bottom or the top to all corners of the nanoparticle, which also results in a change of the nanoparticle performances in different applications. The disappearance of the hybridized plasmonic modes due to the change of the local environment is a dehybridization process.

The out-of-plane high-order modes (γ and β) appear in the spectrum due to the phase variation of the field inside the nanoparticle, along its height (retardation effect). In order to observe the dehybridization process more clearly and without the interference of the high-order modes, we study the SNC of smaller height. The diameter is also reduced to increase the overlap of the spectral positions of the dipolar and quadrupolar resonances.

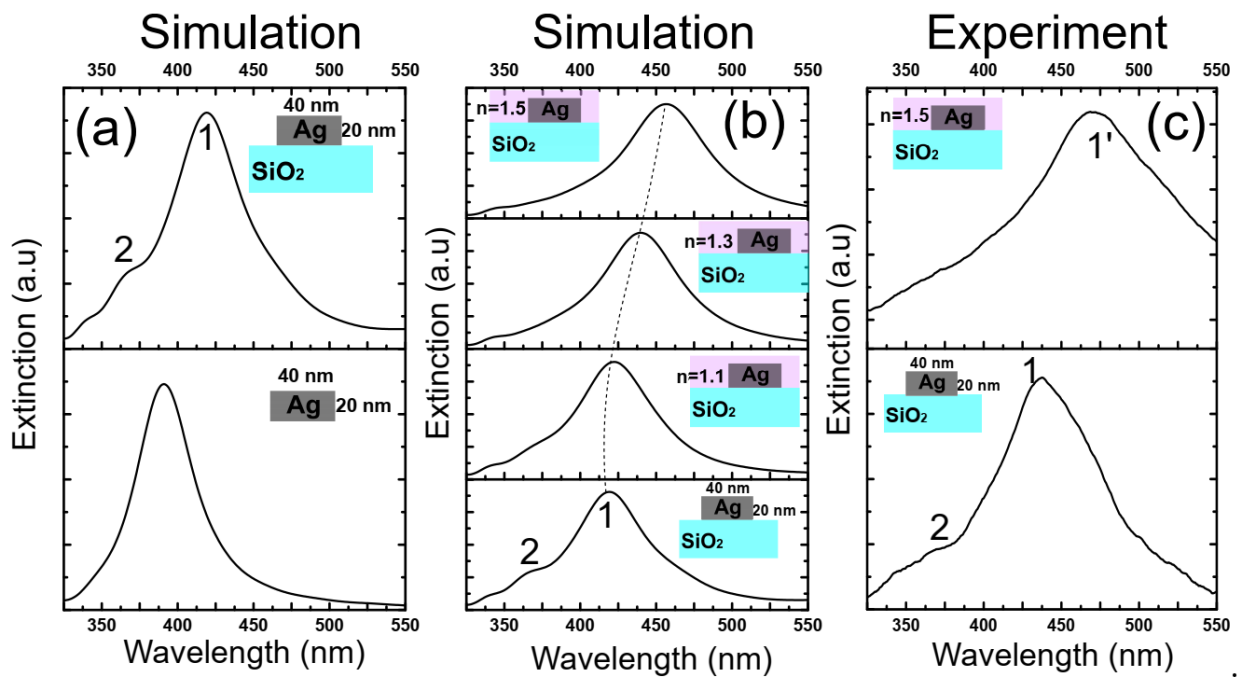

Figure 6: (a) Simulated extinction spectra of a SNC (40 nm-diameter, 20 nm-height) disposed in air and on a glass substrate. (b) Simulated extinction spectra of a SNC deposited on the glass substrate and covered by a layer of refractive index varying from 1.1 to 1.5. (c) Experimental extinction spectra of SNCs (40 nm-diameter, 20 nm-height). The bottom spectrum shows the extinction of SNCs without a superstrate. The upper spectrum shows the extinction of SNCs covered by a PMMA layer. The spectrum of SNCs without the PMMA layer shows two hybridized modes (Mode 1 and Mode 2), while for the covered sample with PMMA show a single dehybridized mode 1'.

In Fig. 6a we depict the calculated extinction spectra of a SNC having a 40 nm-diameter and 20 nm-height. In the bottom spectrum the SNC in air shows an only peak around 390 nm. The SNC deposited on a glass substate already has two excited plasmonic modes (main peak at 418 nm and a shoulder at 370 nm). We denoted the hybridized modes as 1 and 2. Fig. 6b shows the change in the extinction spectra after the addition of the layer with a refractive index higher than air. One can see that with the increase of the refractive index of the superstrate two effects happen: the mode 1 redshifts and the mode 2 diminishes. When there is no layer on top of the nanoparticles the hybridization is present because the contrast of the refractive index between the air and glass is significant and so is the reflection from the glass. In the case of a layer on the nanocylinder, the reflection coming from the glass decreases according to the Fresnel equations. For normal incidence, the reflection coefficient of air/glass interface is $r = (n_{air} - n_{glass})/(n_{air} + n_{glass})$ which is equal to -0.2, while for n=1.3 the r reflection coefficient for superstrate/glass interface drops around 2.8 times to -0.071. Accordingly, the reflectivity drops around 8 times ($R = |r|^2$). Firstly, it results in a weakening of the interference between the incoming and reflected field. Secondly, with the increase of the refractive index, the wavelength of scattered waves in the upper medium is increased. Since the depth of the scattered evanescent field is proportional to the wavelength in the upper medium, the near-field interactions between the substrate and the nanocylinder decrease. The decrease of both near-field and far-field phenomena results in the fall of the quadrupolar mode excitation. Hence, we undo the hybridization by eliminating the quadrupolar mode which was excited due to interference of the incoming and scattered fields.

In Fig. 6c we provide the experimental study of dehybridization of plasmonic modes. The extinction spectrum on the bottom side of Fig. 6c shows two hybridized modes (1 and 2) due to an asymmetrical environment (glass/air). Once we cover the silver nanocylinders with a PMMA layer, which has a similar refractive index to the glass, we observe one nonhybridized plasmonic mode (1'), and the mode (2) vanishes. When we have a homogeneous environment around the nanocylinder and a non-sufficient height to have a retardation effect, we do not observe multipoles excitation in contrast to the case of 50-nm height nanocylinder (Fig. 5). As well we do not observe hybridized modes.

CONCLUSION

In this manuscript we discuss the manifestation of the hybridized modes in the extinction spectrum of the nanoparticle on the substrate. We show experimentally and theoretically that the position of these modes highly depends on the precise environment of the nanoparticles and slight variations of their shape. The modes hybridization happens as a consequence of the quadrupolar mode excitation due to the superposition of the reflected and incoming light, and then it depends on the reflection of the substrate. Hence, when we cover the nanoparticle of interest with any dielectric layer of the refractive index similar to RI of a substrate, it may change the reflectivity of the interface, or, in other terms, the symmetry of the system, and prevent the hybridization of the modes. Moreover, in some cases the increase of the effective refractive index around the nanoparticle results in a higher phase variation inside the nanoparticle, and the multipole modes like quadrupole and hexapole are excited due to the retardation effect, but not due to the hybridization process. These modes have different spatial field distribution and behavior which may influence the performance of the LSPR refractive index sensors. We believe that these findings help to understand the mechanism of the modes hybridization induced by the substrate and, then, will promote the development of the application based on this effect. By using a new approach of decomposition of a nanocylinder in two parts, we visualized the excitation of the quadrupolar mode caused by the phase variations coming from the reflected field from the substrate and the impinging field. Finally, we demonstrated the possibility of plasmonic modes dehybridization by creating small contrast of refractive indexes between the substrate and superstrate.

ACKNOWLEDGEMENTS

This work was prepared in the context of the European COST Action MP1302 NanoSpectroscopy. The authors acknowledge the Nano'Mat platform and their financial support for experimental facilities. The numerical simulations were supported by the HPC Center of Champagne-Ardenne ROMEO. The authors acknowledge the Grand-Est region (Project NanoConv-QuantumPlasm) and FEDER for financial support of the postdoctoral fellowship.

SUPPORTING INFORMATION AVAILABLE

REFERENCES

(1) Giannini, V.; Fernández-Domínguez, A. I.; Heck, S. C.; Maier, S. A. Plasmonic nanoantennas: Fundamentals and their use in controlling the radiative properties of nanoemitters. Chem. Rev. 2011, 111, 3888−3912.

(2) Li, J.-F.; Zhang, Y.-J.; Ding, S.-Y.; Panneerselvam, R.; Tian, Z.-Q. Core-shell nanoparticle-enhanced raman spectroscopy. Chem. Rev. 2017, 117, 5002−5069.

(3) Li, J.-F.; Li, C.-Y.; Aroca, R. F. Plasmon-enhanced fluorescence spectroscopy. Chem. Soc. Rev. 2017, 46, 3962−3979.

(4) Muravitskaya, A.; Rumyantseva, A.; Kostcheev, S.; Dzhagan, V.; Stroyuk, O.; Adam, P.-M. Enhanced Raman scattering of ZnO nanocrystals in the vicinity of gold and silver nanostructured surfaces. Opt. Express 2016, 24, A168.

(5) Wang, J. A review of recent progress in plasmon-assisted nanophotonic devices. Front. Optoelectron. 2014, 7, 320−337.

(6) Mayer, K. M.; Hafner, J. H. Localized surface plasmon resonance sensors. Chem. Rev. 2011, 111, 3828−3857.

(7) Proust, J.; Martin, J.; Gérard, D.; Bijeon, J.-L.; Plain, J. Detecting a Zeptogram ofPyridine with a Hybrid Plasmonic-Photonic Nanosensor. ACS Sens. 2019, 4, 586−594.

(8) Kaminska, I.; Maurer, T.; Nicolas, R.; Renault, M.; Lerond, T.; Salas-Montiel, R.; Herro, Z.; Kazan, M.; Niedziolka-Jönsson, J.; Plain, J.; et al. Near-field and far-field sensitivities of LSPR sensors. J. Chem. Phys. 2015, 119, 9470−9476.

(9) Jatschka, J.; Dathe, A.; Csáki, A.; Fritzsche, W.; Stranik, O. Propagating and localized surface plasmon resonance sensing - A critical comparison based on measurements and theory. Sens. Biosens. Res. 2016, 7,62−70.

(10) Zhang, C.; Lu, Y.; Ni, Y.; Li, M.; Mao, L.; Liu, C.; Zhang, D.; Ming, H.; Wang, P. Plasmonic lasing of nanocavity embedding in metallic nanoantenna array. Nano Lett. 2015, 15, 1382−1387.

(11) Zhou, N.; López-Puente, V.; Wang, Q.; Polavarapu, L.; Pastoriza- Santos, I.; Xu, Q.-H. Plasmon-enhanced light harvesting: Applications in enhanced photocatalysis, photodynamic therapy and photovoltaics. RSC Adv. 2015, 5, 29076−29097.

(12) Besteiro, L. V.; Yu, P.; Wang, Z.; Holleitner, A. W.; Hartland, G. V.; Wiederrecht, G. P.; Govorov, A. O. The fast and the furious: Ultrafast hot electrons in plasmonic metastructures. Size and structure matter. Nano Today 2019, 27, 120−145.

(13) Jiang, N.; Zhuo, X.; Wang, J. Active Plasmonics: Principles, Structures, and Applications. Chem. Rev. 2018, 118, 3054−3099.

(14) Kelly, K. L.; Coronado, E.; Zhao, L. L.; Schatz, G. C. The optical properties of metal nanoparticles: The influence of size, shape, and dielectric environment. J. Chem. Phys. 2003, 107, 668−677.

(15) Krug, M. K.; Reisecker, M.; Hohenau, A.; Ditlbacher, H.; Trügler, A.; Hohenester, U.; Krenn, J. R. Probing plasmonic breathing modes optically. Appl. Phys. Lett. 2014, 105, 171103.

(16) Muravitskaya, A.; Gokarna, A.; Movsesyan, A.; Kostcheev, S.; Rumyantseva, A.; Couteau, C.; Lerondel, G.; Baudrion, A.-L.; Gaponenko, S.; Adam, P.-M. Refractive index mediated plasmon hybridization in an array of aluminium nanoparticles. Nanoscale 2020, 12, 6394−6402.

(17) Movsesyan, A.; Baudrion, A.-L.; Adam, P.-M. Revealing the Hidden Plasmonic Modes ofa Gold Nanocylinder. J. Chem. Phys. 2018, 122, 23651−23658.

(18) Hao, F.; Larsson, E. M.; Ali, T. A.; Sutherland, D. S.; Nordlander, P. Shedding light on dark plasmons in gold nanorings. Chem. Phys. Lett. 2008, 458, 262−266.

(19) Chu, M.-W.; Myroshnychenko, V.; Chen, C. H.; Deng, J.-P.; Mou, C.-Y.; de Abajo, F. J. G. Probing bright and dark surface-plasmon modes in lndividual and coupled noble metal nanoparticles using an electron beam. Nano Lett. 2009, 9, 399−404.

(20) Otte, M. A.; Estévez, M.-C.; Carracosa, L. G.; González- Guerrero, A. B.; Lechuga, L. M.; Sepúlveda, B. Improved biosensing capability with novel suspended nanodisks. J. Chem. Phys. 2011, 115, 5344−5351.

(21) Rycenga, M.; Cobley, C. M.; Zeng, J.; Li, W.; Moran, C. H.; Zhang, Q.; Qin, D.; Xia, Y. Controlling the synthesis and assembly of silver nanostructures for plasmonic applications. Chem. Rev. 2011, 111, 3669−3712.

(22) Lermé, J.; Bonnet, C.; Broyer, M.; Cottancin, E.; Manchon, D.; Pellarin, M. Optical properties of a particle above a dielectric interface: Cross sections, benchmark calculations, and analysis of the intrinsic substrate effects. J. Chem. Phys. 2013, 117, 6383−6398.

(23) Zhang, S.; Bao, K.; Halas, N. J.; Xu, H.; Nordlander, P. Substrate- induced Fano resonances of a plasmonic nanocube: a route to increased-sensitivity localized surface plasmon resonance sensors revealed. Nano Lett. 2011, 11, 1657−1663.

(24) Sherry, L. J.; Chang, S.-H.; Schatz, G. C.; Van Duyne, R. P.; Wiley, B. J.; Xia, Y. Localized surface plasmon resonance spectroscopy of single silver nanocubes. Nano Lett. 2005, 5, 2034−2038.

(25) McMahon, J. M.; Wang, Y.; Sherry, L. J.; Van Duyne, R. P.; Marks, L. D.; Gray, S. K.; Schatz, G. C. Correlating the Structure, Optical Spectra, and Electrodynamics of Single Silver Nanocubes. J. Chem. Phys. 2009, 113, 2731−2735.

(26) Knight, M. W.; Wu, Y.; Lassiter, J. B.; Nordlander, P.; Halas, N. J. Substrates matter: influence of an adjacent dielectric on an individual plasmonic nanoparticle. Nano Lett. 2009, 9, 2188−2192.

(27) Qin, F.; Cui, X.; Ruan, Q.; Lai, Y.; Wang, J.; Ma, H.; Lin, H.-Q. Role of shape in substrate-induced plasmonic shift and mode uncovering on gold nanocrystals. Nanoscale 2016, 8, 17645−17657.

(28) Dmitriev, A.; Hägglund, C.; Chen, S.; Fredriksson, H.; Pakizeh, T.; Käll, M.; Sutherland, D. S. Enhanced nanoplasmonic optical sensors with reduced substrate effect. Nano Lett. 2008, 8, 3893−3898.

(29) Martinsson, E.; Otte, M. A.; Shahjamali, M. M.; Sepulveda, B.; Aili, D. Substrate effect on the refractive index sensitivity of silver nanoparticles. J. Chem. Phys. 2014, 118, 24680−24687.

(30) Mahmoud, M. A.; Chamanzar, M.; Adibi, A.; El-Sayed, M. A. Effect of the dielectric constant of the surrounding medium and the substrate on the surface plasmon resonance spectrum and sensitivity factors ofhighly symmetric systems: Silver nanocubes. J. Am. Chem. Soc. 2012, 134, 6434−6442.

(31) Bottomley, A.; Prezgot, D.; Ianoul, A. Plasmonic properties of silver nanocube monolayers on high refractive index substrates. Appl. Phys. A: Mater. Sci. Process. 2012, 109, 869−872.

(32) Abdijalilov, K.; Schneider, J. B. Analytic field propagation TFSF boundary for FDTD problems involving planar interfaces: Lossy material and evanescent fields. IEEE Antenn. Wireless Propag. Lett. 2006, 5, 454−458.

(33) Gedney, S. D.; Zhao, B. An auxiliary differential equation formulation for the complex-frequency shifted PML. IEEE Antenn. Wireless Propag. Lett. 2010, 58, 838−847.

(34) Hagemann, H.-J.; Gudat, W.; Kunz, C. Optical constants from the far infrared to the x-ray region: Mg, Al, Cu, Ag, Au, Bi, C, and Al2O3. J. Opt. Soc. Am. 1975, 65, 742−744.

(35) Schmidt, F.-P.; Ditlbacher, H.; Hohenester, U.; Hohenau, A.; Hofer, F.; Krenn, J. R. Dark plasmonic breathing modes in silver nanodisks. Nano Lett. 2012, 12, 5780−5783.

(36) Chakrabarty, A.; Wang, F.; Minkowski, F.; Sun, K.; Wei, Q.-H. Cavity modes and their excitations in elliptical plasmonic patch nanoantennas. Opt. Express 2012, 20, 11615.

(37) Dai, F.; Horrer, A.; Adam, P. M.; Fleischer, M. Accessing the Hotspots of Cavity Plasmon Modes in Vertical Metal-Insulator-Metal Structures for Surface Enhanced Raman Scattering. Adv. Opt. Mater. 2020, 8, 1901734.

(38) Lamri, G.; Movsesyan, A.; Figueiras, E.; Nieder, J. B.; Aubard, J.; Adam, P.-M.; Couteau, C.; Felidj, N.; Baudrion, A.-L. Photochromic control of a plasmon-quantum dots coupled system. Nanoscale 2019, 11, 258−265.

(39) Movsesyan, A.; Marguet, S.; Muravitskaya, A.; Béal, J.; Adam, P.-M.; Baudrion, A.-L. Influence of the CTAB surfactant layer on optical properties of single metallic nanospheres. J. Opt. Soc. Am. A 2019, 36, C78.